\documentstyle[12pt]{article}
\begin{document}

\begin{center}
{\large \bf Memory Effect in Silicon Nitride in Silicon Devices}

\end{center}

\begin{center}
{V. A. Gritsenko$^*$, Yu. N. Morokov$^{a}$, Yu. N. Novikov, and J. B. Xu$^{b}$}
\end{center}

Institute of Semiconductor Physics, Novosibirsk, 630090, Russia \\
$^{a}$ Institute of Computational Technologies, Novosibirsk, 630090, Russia \\
$^{a}$Electronic Engineering Department, The Chinese University of Hong Kong,\\
Shatin, N.T., Hong Kong
 
\begin{abstract}
{Amorphous silicon oxide (a-SiO$_2$) and nitride (a-Si$_3$N$_4$) are two key 
dielectrics in microelectronic silicon devices [1]. The dominant dielectric 
used currently in silicon devices is a-SiO$_2$ [1,2]. Application of silicon 
oxide for future devices will be impeded by several fundamental limitations 
which lead to low reliability of semiconductor devices and to the necessity of 
alternative dielectrics [1-3]. Amorphous silicon nitride and oxynitride are 
considered now as alternative to a-SiO$_2$ in future devices [4]. One of the 
unique property of a-Si$_3$N$_4$ in comparison with a-SiO$_2$ is the electron 
and hole capture by the deep traps with extremely long life time (10 years) 
in the captured state (the memory effect). This property is widely employed in 
memory devices and microprocessors in computers. Despite numerous efforts the 
nature of traps responsible for the memory effect in Si$_3$N$_4$ is so far 
unclear. In this paper we discuss the nature of such traps using the 
quantum-chemical simulation. The calculations show that the defects 
responsible for the electron and hole capture can be the Si-Si defects created 
by excess silicon atoms in Si$_3$N$_4$.} 
\end{abstract}

Silicon devices in integrated circuits are the basic components of modern 
electronics and information technology. Silicon electronics has globally 
changed our life and will have great impacts in future. Currently the 
a-SiO$_2$ films of about 5 nm thickness are used as a gate dielectric in 
silicon devices. The dielectric films with a thickness of 1.5-2.0 nm are 
required for future 1 Gbite devices [1]. The a-SiO$_2$ with such thickness has 
low reliability and therefore cannot be applicable [1,2]. 

Unlike a-SiO$_2$ that exists widely in the nature, amorphous silicon nitride 
has been specially synthesised for microelectronic applications and has 
several advantages over a-SiO$_2$. Silicon nitride is used as insulator in 
silicon devices, as a mask against the impurity diffusion during fabrication, 
and as a passivation layer to protect the devices against the environment. 
Intrinsically a-Si$_3$N$_4$ contains a high density of the electron and hole 
traps and therefore is widely used in triple oxide-nitride-oxide (ONO) 
structures in silicon devices [5,6]. The stacked ONO structures are used as 
dielectric in memory capacitors of memory silicon devices and in 
microprocessors that are the main elements of personal computers. The 
advantage of the ONO dielectric over a-SiO$_2$ is the high reliability due to 
the electron capture in Si$_3$N$_4$ [6]. This effect results in the increase 
of a breakdown field strength and the improved reliability of devices composed 
of ONO dielectrics. Besides, the ONO structures are the key components in 
silicon memory transistors based on the electron and hole capture in 
Si$_3$N$_4$ [7-9]. Information written in the transistor is stored and saved 
without the energy consumption more than ten years (similar to a magnetic 
memory). According to projections this type of the silicon memory devices will 
replace the floppy and hard magnetic discs in computers in future. 

The memory effect in a-Si$_3$N$_4$ was discovered in 1968, in the beginning of 
silicon microelectronics development [10]. As early as in 1994, the Hitachi 
Corporation has designed the 1 Mbit memory array based on the memory effect 
in a-Si$_3$N$_4$ [8], and Sony Corporation has designed memory devices for 
future 256 Mbit [9]. Furthermore, one of the unique advantages of 
a-Si$_3$N$_4$ over a-SiO$_2$ is a high tolerance to radiation, in particular, 
to cosmic rays. Therefore, silicon devices, based on a-Si$_3$N$_4$, are widely 
used in the space and nuclear technology with the harsh radiation environment 
[11]. 

The nature (atomic structure) of traps responsible for the memory effect in 
a-Si$_3$N$_4$ is so far unclear in spite of the extremely technological 
importance, numerous applications, and studies. However, there is a consensus 
that the capture of electrons and holes is related with the defects created by 
excess silicon atoms in a-Si$_3$N$_4$ [12,13]. 

The first idea was to associate the memory effect with the existence of the 
neutral three-fold coordinated silicon atoms $\equiv$Si* with unpaired 
electron (K$^0$ center) which can capture either electron or hole. However, a 
very important physical information about the capturing properties of traps in 
a-Si$_3$N$_4$ was obtained by electron spin resonance (ESR) experiments, which 
allowed to observe the defects with unpaired electrons. It was found that the 
ESR signal was absent in as-prepared Si$_3$N$_4$ samples [12,13]. This means 
that there are no defects associated with unpaired electrons in as synthesed 
a-Si$_3$N$_4$. Consequently, the neutral isolated $\equiv$Si* defects are 
absent in this materal. 

The next step was to assume that a pair of the three-fold coordinated 
$\equiv$Si* neutral defects creates a pair of the negatively charged 
$\equiv$Si: and positively charged $\equiv$Si defects (K$^-$ and K$^+$ 
centers) with the reaction [12,14,15] 
\begin{eqnarray}
\label{1}
\equiv{\rm Si*} \; + \; \equiv{\rm Si*} \enskip \rightarrow \enskip 
\equiv{\rm Si}:  \; +  \; \equiv{\rm Si}. 
\end{eqnarray}
This model is called the "negative correlation energy model" (NCE) that means 
that the reaction (1) is energetically favorable [12,14,15]. 

The NCE model explains the absence of the ESR signal in as synthesed 
a-Si$_3$N$_4$ because the charged defects $\equiv$Si and $\equiv$Si: do not 
contain unpaired electrons. The NCE model assumes that the reaction (1) may be 
favourable due to a lattice relaxation. According to this model the positively 
charged $\equiv$Si defect is an electron trap and the negatively charged 
$\equiv$Si: defect is a hole trap. 

Another model for interpretation of the memory effect in silicon nitride was 
considered in [7,13,16,17]. That model assumes the existence of the neutral 
Si-Si bonds in a-Si$_3$N$_4$ that are responsible for the electron or hole 
capture. The Si-Si defect model explains the absence of the ESR signal in 
as-prepared Si$_3$N$_4$ because two electrons of the neutral Si-Si bond are 
paired. 

The validity of the NCE and Si-Si models is so far under discussion [18]. 

We considered these models with the quantum-chemical MINDO/3 method which we 
applied early for the defect simulation [19]. Previously this method was used 
also for calculation of the electronic structure of the $\equiv$Si* and 
$\equiv$Si-Si$\equiv$ defects in a-SiO$_2$ [20]. 

The 84 atomic (Si$_{20}$N$_{28}$H$_{36}$) and 70 atomic 
(Si$_{18}$N$_{17}$H$_{35}$) clusters were used for simulation of the 
a-Si$_3$N$_4$ bulk electronic structure to estimate the energy gain with 
capture of an electron or hole. The 56 atomic (Si$_{14}$N$_{12}$H$_{30}$) and 
the 55 atomic (Si$_{10}$N$_{18}$H$_{27}$) clusters were used for simulation of 
the $\equiv$Si-Si$\equiv$ and $\equiv$Si* defects. To understand the capturing 
properties of defects we added one electron or hole to the cluster. Atomic 
relaxation of the atoms in the first coordination sphere of defects was 
considered in the simulation. 

To verify if the reaction (1) is energetically favorable or not we calculated 
the difference between the total energies of the pairs of clusters ($\equiv$Si 
+ :Si$\equiv$) and ($\equiv$Si* + $\equiv$Si*). Calculations give the positive 
value for this difference $\approx$ 4.0 eV. Ab initio calculations at the 
6-3G*/MP2 level for the cluster Si(NH$_2$)$_3$ give $\approx$ 5.5 eV [21]. 
Taking into account the energy of the long range polarization (estimated with 
"classical model" [18]) we obtain the next values for the total energy gain 
$\Delta$E $\approx$ 2.7 eV (MINDO/3) and $\Delta$E $\approx$ 3.5 eV (ab 
initio).  

The recent first principle DFT calculations of the $\equiv$Si* and 
$\equiv$ Si$_2$N* defects in silicon nitride [18] gave a positive correlation 
energy 0.9 eV for the reaction (1) with isolated $\equiv$Si* defects. The 
small value of $\Delta$E is explained by a strong relaxation effect observed 
in [18] for the positively charged defect $\equiv$Si. 

The positive value of $\Delta$E means that the charged $\equiv$Si and 
$\equiv$Si: states of a three-fold coordinated silicon atom are energetically 
unfavorable in comparison with the neutral defects $\equiv$Si*. We may 
conclude that the calculations do not support for the time being the widely 
accepted NCE model for a-Si$_3$N$_4$. 

To understand the capturing properties of the Si-Si bond we studied this 
defect in different charge states. The calculations show that the energy gain 
with capture of a hole by the Si-Si defect is about 0.34 eV. The capture of an 
electron is accompanied by the energy gain of 1.76 eV. Hence, the Si-Si bond 
in a-Si$_3$N$_4$ can capture either electron or hole. The absence of the ESR 
signal in a-Si$_3$N$_4$ with captured electrons or holes is explained by the 
interaction of electrons (holes) captured in the nearest occupied Si-Si 
defects [17]. According to our consideration the model of the Si-Si bond is 
the most favorable for explanation of the memory effect in a-Si$_3$N$_4$. 

\newpage

\begin{center}
{\bf References}
\end{center}

$^*$ Electronic address: grits@isp.nsc.ru \\
1.  Semiconductor Industry Association. {\it International Technology 
    Roadmap for Semiconductors: 1999 edition.} (Austin, TX:International 
    SEMATECH, 1999); http://www.sematech.org. \\
2.  D.A. Muller, T. Sorsch, S. Moccio, F.H. Baumann, K. Evans-Lutterodt, and 
    G. Timp, Nature {\bf 399} (1999) 758. \\
3.  M. Shulz, Nature {\bf 399} (1999) 729. \\
4.  Y. Shi, X. Wang, and T.P. Ma, IEEE Transactions on Electron Devices 
    {\bf 46} (1999) 362. \\
5.  S. Kimura and H. Ikoma, J. Appl. Phys. {\bf 85} (1999) 551. \\
6.  V.A. Gritsenko, In {\it Proc. of NATO Advanced Research Workshop: 
    Fundamental Aspects of Ultrathin Dielectrics on Si-based  Devices toward 
    an Atomic Scale Understanding. High Technology}. Vol. {\bf 47} 
    (Klumer Academic, Boston, USA, 1998) 335. \\
7.  V.A. Gritsenko, in {\it Silicon Nitride in Electronics}, A.V. Rzhanov, Ed. 
    (Elsevier, New York, 1988). \\
8.  S. Minami, K. Ujiie, M. Terasawa, K. Komori, K. Furusawa, and Y. Kamigaki, 
    IEICE Transactions on Electronics {\bf E77C} (1994) 1260. \\
9.  T. Bom, A. Nakamura, H. Aozasa, M. Yamagishi, and Y. Komatsu, 
    Jap. J. Appl. Phys. {\bf 35} (1996) 898. \\
10. H.C. Pao and M. O'Connel, Appl. Phys. Lett. {\bf 12} (1968) 260. \\
11. J.R. Schwank, S.B. Roeske, D.E. Beutler, D.J. Moreno, and A. Shaneyfelt, 
    IEEE Transaction on Nuclear Science {\bf 43} (1996) 2671. \\
12. W.L. Warren, J. Kanicki, J. Robertson, E.H. Poindexter, and 
    P.J. McWhorter, J. Appl. Phys. {\bf 74} (1993) 4034. \\
13. V.A. Gritsenko and A.D. Milov, JETP Lett. {\bf 64} (1996) 531. \\
14. P.W. Anderson, Phys. Rev. Lett. {\bf 34} (1975) 953. \\
15. R.A. Street and N.F. Mott, Phys. Rev. Lett. {\bf 35} (1975) 1293. \\
16. V.A. Gritsenko, J.B. Xu, R.W.M. Kwok, I.Y. Ng, and I.H. Wilson, 
    Phys. Rev. Lett. {\bf 35} (1998) 1054. \\
17. V.A. Gritsenko, JETP Lett. {\bf 64} (1996) 525. \\
18. G. Pacchioni and D. Erbetta, Phys. Rev. {\bf B61} (2000) 15005. \\
19. V.A. Gritsenko, P.M. Lenahan, Yu.N. Morokov, and Yu.N. Novikov, "Nature of 
    traps responsible for failure of MOS devices", cond-mat/0011241. \\
20. A.X. Chu and W.B. Fowler, Phys. Rev. {\bf B41} (1990) 5061. \\
21. V.A. Gritsenko, Yu.N. Morokov, Yu.N. Novikov, S.P. Pang, R.M.W. Kwok, 
    Z.F. Liu, and L.W.M. Lau, unpublished. 

\end{document}